\documentstyle[twocolumn,aps]{revtex}

\newcommand{\kv}{\mbox{{\bf k}}}
\newcommand{\pv}{\mbox{{\bf p}}}
\newcommand{\Av}{\mbox{{\bf A}}}
\newcommand{\Hv}{\mbox{{\bf H}}}
\newcommand{\Rv}{\mbox{{\bf R}}}

\newcommand{\lv}{\mbox{{\bf l}}}

\newcommand{\ev}{\mbox{{\bf e}}}
\input{epsf}
\begin{document}
\draft
\title{Upper critical field for electrons in two--dimensional lattice}
\author{Marcin Mierzejewski and Maciej M. Ma{\'s}ka 
\thanks{e-mail: maciek@risc3.phys.us.edu.pl }}
\address{Department of Theoretical Physics, University of Silesia, 40-007 Katowice,
Poland}
\maketitle
\begin{abstract}
We address a problem of the upper critical field in a lattice described
by a two--dimensional tight--binding model with the on--site pairing. 
We develop a finite--system--approach which enables investigation of 
magnetic and superconducting properties of electrons on clusters,
consisting of a few thousand sites. We discuss how the quasiparticle
density of states changes with the applied external magnetic field 
and present the temperature dependence of the upper critical field.   
We also briefly discuss possible extension of the model 
to account for the properties of high--temperature superconductors.
\end{abstract}

The issue of the critical field consists of two different phenomena,
namely a movement of electrons in a periodic potential under influence 
of a magnetic field and superconductivity. Each of these phenomena 
has been
investigated since many decades and many solutions in limiting 
cases are known at present. Concerning the movement of 
electrons one deals with two limiting cases: free or nearly free 
electrons in a magnetic field, when Landau levels structure sets on
 \cite{gorkov,helf}, and electrons in a periodic potential in the 
absence of 
magnetic field, when the solutions are Bloch waves which lead to energy bands.  
Away from these limiting cases the situation is much more complicated. 
Application of magnetic field to the two--dimensional (2D) electron system
in tight--binding
approximation leads to a fractal energy spectrum known as Hofstadter's 
butterfly, where very small changes in magnetic field can
result in a completely different spectrum \cite{harper,lang,hof}. 
%
Electrons on a lattice are gauge--invariantly coupled with a U(1) gauge
field by introducing phase factors in the kinetic--energy hopping term,
i.e., the wave function acquires a factor 
$\exp\left(\frac{ie}{\hbar c}\int^j_i {\rm {\bf A}} \cdot d{\rm {\bf l}}
\right)$, where {\bf A} is the external classical vector potential, 
when  an electron hops from site $i$ do site $j$ \cite{peierls}. 
The Zeeman term is neglected. The same energy spectrum can be obtained in 
a nearly--free--electron method  with a weak periodic perturbation
introduced into the Landau--quantized 2D electron system 
\cite{lang,thoul,usov}.

On the other hand, the influence of a magnetic field on superconductivity
is usually described by phenomenological Ginzburg--Landau theory \cite{GL},
(or Lawrence--Doniach theory in the case of layered superconductors \cite{LD})
where the magnetic field is treated semiclassically. This approach was
later justified also at the microscopic level \cite{gorkov}, but 
the temperature dependence of physical quantities 
is also of a phenomenological character
and therefore its validity is limited.   

Although, there is a general agreement that 
external magnetic field reduces the
critical temperature, positive curvature of $H_{c2}\left(T\right)$ observed
in high--temperature
superconductors \cite{hc2} is still a matter at issue. 
The most important differences between standard BCS--type and high--temperature
superconductors are related to the presence of strong 
electronic correlations and
specific geometry of high--$T_{c}$ materials.     
In this paper we address an important problem concerning 
the upper critical field for electrons described by the two--dimensional
tight--binding model.

Our starting point is two-dimensional square lattice
immersed in a perpendicular, uniform magnetic field. The 
mean-field Hamiltonian is of the form
\begin{equation}
\hat{H}=\sum^{}_{i,j,\sigma} t^{}_{ij}\left(\Av \right) 
c^{\dagger}_{i\sigma}c^{}_{j\sigma}\:-\:
V\sum^{}_{i}\left(c^{\dagger}_{i\uparrow}c^{\dagger}_{i\downarrow}\Delta^{}_{i}
+c^{}_{i\downarrow}c^{}_{i\uparrow} \Delta^{\star}_{i} \right).
\end{equation}
Here, $c^{\dagger}_{i\sigma}$ ($c^{}_{i\sigma}$) creates (annihilates)
an electron with spin $ \sigma$ on site $i$, $V$ stands for the magnitude
of the on-site attraction and $\Av$ is the vector potential corresponding 
to the external magnetic field $\Hv$. 
Similarly to Gor'kov's approach we introduce local superconducting order
parameter \cite{gorkov}
\begin{equation}
\Delta^{}_{i}=\left<c^{}_{i\downarrow}c^{}_{i\uparrow}\right>,
\end{equation}
which, in general, can be site--dependent.
According to the Peierls substitution \cite{peierls} the original
hopping integral $t_{ij}$ is multiplied by a phase factor, which accounts for
coupling of electrons to the magnetic field
\begin{equation}
t^{}_{ij}\left(\Av \right)= t^{}_{ij}
\exp\left(\frac{ie}{\hbar c} \int^{\Rv_{i}}_{\Rv_{j}}
\Av\cdot d\lv\right). 
\end{equation}
In order to derive the self-consistent equation for the gap-function, we
make use of unitary transformation and introduce 
\begin{equation}
a^{}_{m\sigma}=\sum^{}_{i}   U^{\dagger}_{mi} c^{}_{i \sigma},
\end{equation}
where the unitary matrix $U^{}_{im}$ consists of eigenvectors of the
hermitian matrix $t^{}_{ij}\left(\Av\right)$ 
\begin{equation}
\sum^{}_{i,j}U^{\dagger}_{mi}
t^{}_{ij}\left(\Av\right) U^{}_{jn}=\delta^{}_{mn}E^{}_{m}
\end{equation}
This unitary transformation determines energy spectrum of the system in the 
normal state in the presence of external magnetic field.
In the absence of magnetic field, Eq. (4) represents a transformation
to the momentum space, namely $U_{jm}=1/N \exp\left(i \Rv_{j}\cdot\kv_{m} \right)$.
In the case of free electron gas external magnetic field leads 
to the occurrence of rotationally invariant states corresponding to the Landau orbits. 
In order to visualize the impact of magnetic field on electrons in 
the 2D lattice we have calculated the resulting current distribution.
Within the framework of linear--response theory the current operator 
is given by $\hat{J}_l(x,y)=-\partial \hat{H}/\partial A_l(x,y)$, 
where $(x,y)$ denotes spatial coordinates and $l$'s are unit 
vectors in the lattice axes directions. Results obtained on a finite 
cluster with applied magnetic field 
$e a^{2}H^{}_{z}/(\hbar c)=0.1$  are presented in Figure 1. Note 
that radiuses of corresponding Landau orbits are of the order of a few
lattice constants $a$.        
\vspace{9 cm}
\begin{figure}[h]
\includegraphics{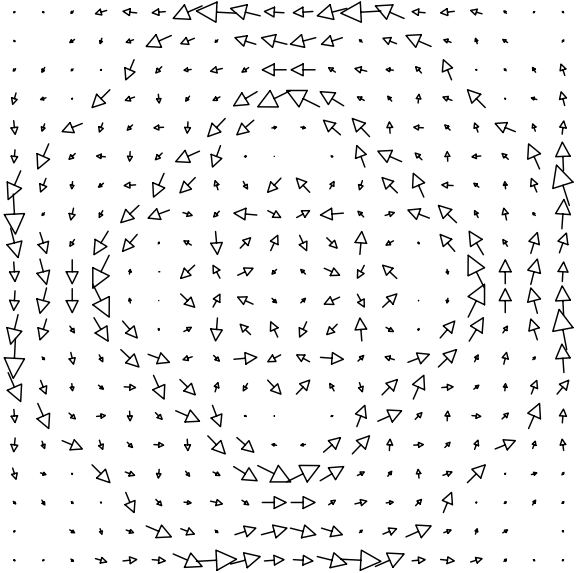}
\caption{Current distribution originating from external magnetic field
calculated for a 20$\times$20 cluster with fixed boundary conditions. 
Only nearest-neighbor hopping has been taken into
account.
 The size of an arrow is proportional to the
magnitude of current. }
\end{figure}

Equations of motion lead to formally exact expression for the anomalous
Green's function
\begin{eqnarray}
\left(\omega-E^{}_{m}\right)
\left<\left<a^{}_{m\uparrow}\mid a^{}_{n\downarrow} \right>\right>
&=& -V \sum^{}_{i,n^{\prime}_{}} \Delta^{}_{i}
U^{\star}_{im} U^{\star}_{in^{\prime}_{}} \nonumber \\
&& \times 
\left<\left<a^{\dagger}_{n^{\prime}_{}\uparrow}\mid a^{}_{n\downarrow}
\right>\right>.
\end{eqnarray}

At temperatures close to $T^{}_{c}$ one can restrict himself only to
terms linear in the superconducting order parameter.
Therefore, the Green's function 
$\left<\left<a^{\dagger}_{n^{\prime}_{}\uparrow}\mid a^{}_{n\downarrow}
\right>\right> $ which enters the above equation 
can be calculated in the normal state and 
one can derive a system of linearized gap equations which determine 
the upper critical field
\begin{eqnarray}
\Delta^{}_{j} &=& V \sum^{}_{i} \Delta^{}_{i}
\sum^{}_{m,n} U^{}_{jn}U^{}_{jm}U^{\star}_{im}U^{\star}_{in}
\nonumber \\
&& \times
\frac{\tanh\displaystyle\frac{E_m}{2k^{}_{B}T}+
      \tanh\displaystyle\frac{E_n}{2k^{}_{B}T}
}{2\left(E^{}_{m}+E^{}_{n}\right)}
\end{eqnarray}
In the absence of
magnetic field one can choose the order parameter as a site-independent
quantity
($\Delta_{i} = \Delta$). Then the above equation can be easily reduced to 
the standard BCS form.
This formula is gauge-invariant and is valid for any dispersion relation
determined  by the hopping integral $t_{ij}$. 
However, in order to simplify numerical calculations we restrict ourselves
only to the nearest-neighbor hopping with $t_{\left<i,j\right>}=t$
and choose the Landau gauge 
$\Av=H^{}_{z}\left(0,x,0\right)$.
Here, $H_{z}$ denotes a magnitude of external magnetic
field perpendicular to the $(x,y)$ plane. Such a form of $\Av$
neglects the effects of diamagnetic screening supercurrents 
induced by the applied field  what means that the true vector
potential should be determined self-consistently.
However, close to the transition temperature i.e.,
in the limit of infinite London penetration depth, such an
approximation is correct.   
Then, the unitary matrix $U^{}_{in}$ takes on the form:
\begin{equation}
U_{in} \equiv U_{x,y}\left(\bar{p}_{x},p_{y}\right)=
N^{-1/4}\;e^{\displaystyle ip^{}_{y}ya}
g\left(\bar{p}^{}_{x},p^{}_{y},x\right),
\end{equation}   
where $\left(x,y\right)$ enumerate the lattice sites 
$\Rv^{}_{x,y}=\ev^{}_{x} a x+\ev^{}_{y} a y$  and
$p^{}_{y}$ is the wave vector in the $y$ direction.  

It follows from
Eq. (5) that $g\left(\bar{p}^{}_{x},p^{}_{y},x\right)$ must fulfill
the Harper's equation \cite{harper}:
\begin{eqnarray}
&& g\left(\bar{p}^{}_{x},p^{}_{y},x+1\right)+
2\cos\left(h x -p^{}_{y}a\right)g\left(\bar{p}^{}_{x},p^{}_{y},x\right)
\nonumber \\
&&+ g\left(\bar{p}^{}_{x},p^{}_{y},x-1\right)=
t^{-1}_{}E\left(\bar{p}^{}_{x},p^{}_{y}\right)
g\left(\bar{p}^{}_{x},p^{}_{y},x\right),
\end{eqnarray}
where
$
h=e a^{2}H^{}_{z}/(\hbar c)
$.
$h/(2 \pi)$ can be interpreted as a ratio of the 
flux through a lattice cell to one flux
quantum \cite{hof}. In the absence of magnetic field $\bar{p}^{}_{x}$ corresponds to the
$x$ component of the wave vector $\pv$. As the choice of Landau gauge breaks
the translational symmetry along the $x$ axis, for  $\Hv \neq {\bf 0}$  
$\bar{p}^{}_{x}$ represents a quantum number which, however, can not be
identified as a component of the wave vector.  The choice of this
gauge  allows one to take $U_{in}$ in the form given by Eq. (8) and reduces
the original two-dimensional eigenproblem (diagonalization of the kinetic
part of the Hamiltonian)
to a one--dimensional difference equation. A thorough analysis of
Harper's equations can be found in Ref. \cite{hof}.  

Due to the plane-wave behavior in the $y$ direction, there is a solution of
Eq.(7) which does not depend explicitly on $y$:
\begin{equation}
\Delta^{}_{i}\equiv\Delta^{}_{x,y}=\Delta^{}_{x}
\end{equation}
For the specific form of $U_{in}$, as given by Eq. (8),  
the upper critical field is determined by the following equation:
\begin{eqnarray}
\Delta^{}_{x^{\prime}_{}}&=&\frac{V}{\sqrt{N}}\sum^{}_{x} \Delta^{}_{x}
\sum_{p^{}_{y},\bar{p}^{}_{x},\bar{k}^{}_{x}}
g\left(\bar{p}^{}_{x},p^{}_{y},x^{\prime}_{}\right) 
g\left(\bar{k}^{}_{x},-p^{}_{y},x^{\prime}_{}\right) \nonumber \\
& & \times \; g\left(\bar{k}^{}_{x},-p^{}_{y},x\right) 
g\left(\bar{p}^{}_{x},p^{}_{y},x\right) \nonumber \\
& & \times \; \frac{\tanh\displaystyle\frac{E\left(\bar{p}^{}_{x},
p^{}_{y}\right)}{2k^{}_{B}T}
+\tanh\displaystyle\frac{E\left(\bar{k}^{}_{x},-p^{}_{y}
\right) }{2k^{}_{B}T}
}{2\left[E\left(\bar{p}^{}_{x},p^{}_{y}\right)+E\left(\bar{k}^{}_{x},-
p^{}_{y}\right) \right]}.
\end{eqnarray}

In order to evaluate the upper critical field one has to start with solving the
Harper's equation. The corresponding energy spectrum
$E\left(\bar{p}_{x},p_{y} \right)$ has been obtained for the first time by
Hofstadter and constitutes a self--similar, fractal structure  
known as the Hofstadter butterfly \cite{hof}. 

In the thermodynamic limit Eq. (11) is actually an infinite system of equations and
can not be solved exactly. Therefore, in order to get the first insight into 
the properties of the system under consideration, we have performed numerical
calculations for finite systems. In particular, we have investigated square
clusters which consist of a few thousand  lattice sites. Due to the broken
translational symmetry, we have introduced periodic
boundary conditions only  along the $y$ axis and  fixed boundary conditions
in $x$ direction. It means that for the $M \times M$ cluster  
we have taken into
account $M$ points along $x$ axis and $M$ values of the wave vector
$p_{y} \in \left(-\pi/a,\pi/a\right] $. 
\vspace{7.3 cm}
\begin{figure}[h]
\includegraphics{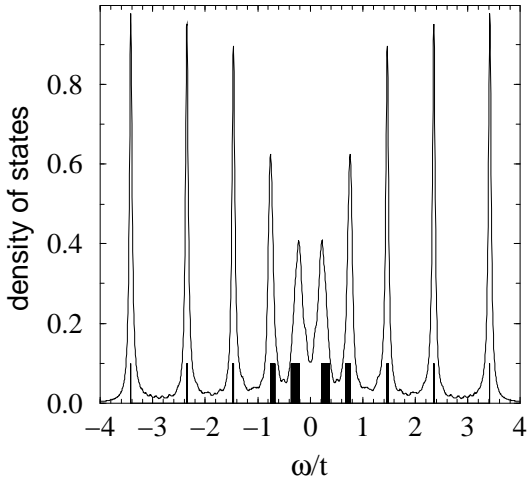}
\caption{Density of states obtained for a $50 \times 50$ cluster and
$h=2 \pi /10$.  
Vertical bars show the energy spectrum obtained for infinite system
(Hofstadter's butterfly). Energy levels obtained from the cluster
calculations are
represented by the Lorentz function with the width equal $0.03t$.}
\end{figure}

\noindent
Making use of the fixed 
boundary conditions in the $x$ direction the Harper equation (9) 
simplifies to an eigenproblem of a tridiagonal  matrix
with all the off-diagonal elements equal unity.
An additional effect originating from such specific boundary
conditions is the absence of unphysical degeneracy of states 
at the Fermi level which occurs in cluster calculations with
fixed and periodic boundary conditions taken in both directions
\cite{yoko}.

To check the influence of finite size effects we have calculated the density
of states of 2D electron  lattice gas in the normal state.
Fig. 2 shows that the density of states calculated for a $50 \times 50$ cluster 
reproduces very well the  results obtained on the basis 
of Hofstadter's procedure for infinite system \cite{hof}.

As a further verification of our cluster approach we have compared
the critical temperature calculated without the external magnetic field
with exact results for the 2D lattice obtained from the BCS-type gap equation
(see the inset in Fig. 3). The critical temperature obtained from 
the cluster calculations is always a bit lower than the exact value simply due 
to the absence of van Hove singularity in finite cluster. 
Since our method works well 
in both limiting cases i.e., in the normal state influenced by external
magnetic field and in superconducting state analyzed without the field,
we  have used this approach to tackle the problem 
which is fundamental in the intermediate region, namely
the influence of the external field on superconductivity.
Numerical solutions of Eq. (11) are shown in Fig. 3. 
 \vspace{7.3 cm}
\begin{figure}[h]
\includegraphics{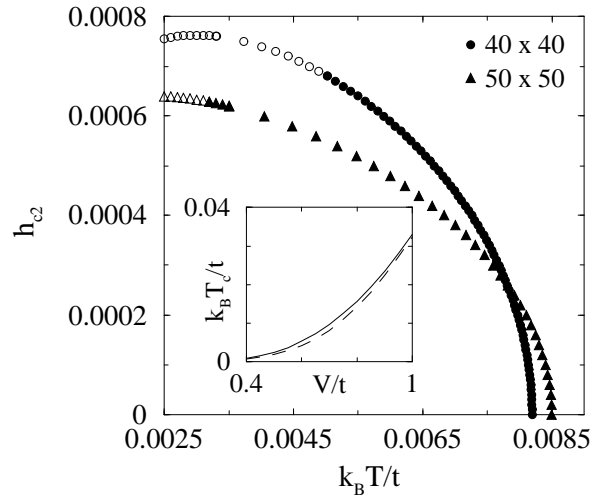}
\caption{Reduced upper critical field obtained for 
$40 \times 40$ and $50 \times 50$ clusters 
plotted as a function of temperature. $V=0.7t$ has been used.
Filled circles and triangles denote results which fulfill the criterion
$k_{B} T > 8t/M^2$. The inset shows the superconducting transition
temperature calculated in the absence of magnetic field versus
the magnitude of pairing potential. Here, continuous and dashed lines
show  exact result for infinite lattice and 
solution for $40 \times 40 $ cluster,
respectively. \\}  
\end{figure}

However, one has to bear in mind that our cluster calculations are not valid
in genuinely low temperatures when the Cooper pair susceptibility 
accounts only  for very few poles of the Green's function
[$E\left(\bar{p}_x,p_y\right)$]  instead of a continuous density of states.   
The simplest criterion of validity for a $ M \times M$ cluster  is an
assumption that temperature ($k_{B}T$) must be larger than an average distance
between different quasiparticle energies ($\simeq 8t/M^2 $).  
We have found that, similarly to the free electron gas described by Gor'kov equations 
\cite{gorkov,helf}, also in the tight--binding model
$H_{c2}\left(T\right)$  exhibits the negative curvature.  
It is remarkable that the critical temperature 
is a smooth function of applied magnetic field 
despite the fact that the energy spectrum is strongly affected even by
small changes of the field \cite{hof}. However, for realistic values of 
magnetic field, $h \sim 10^{-4} \div 10^{-3} $, the band splits into a huge number of
subbands ($\propto h^{-1}$) with the gaps between them  much smaller than $k_{B}T$. 
Therefore, macroscopic properties of the system remain smooth functions
of the magnetic field.

To conclude, we have investigated the relationship between 
superconductivity and an external magnetic field in a simple model
of electron gas in a 2D lattice. The proposed method allows one to analyze
such a model on large clusters, of size of the order of a few thousand
lattice sites, carrying out exact calculations. The main result is the temperature
dependence of the upper critical field (or, equivalently, the applied
magnetic field dependence of the critical temperature). We have found 
that the external magnetic field suppresses 
superconductivity, what remains in agreement with a general filling,
although there are papers where the reentrance of the superconductivity
due to Landau quantization in strong magnetic field is analyzed \cite{tesan}.      
In contradistinction to the Gor'kov approach, which is valid for free
electron gas, we have
carried out calculations for a system with different geometry i.e.,
for the 2D electron gas on a square lattice. Such a geometry is believed
to play a crucial role in high--critical--temperature superconductors.
However, in spite of the differences between these two approaches the
curvature of the transition curve $H_{c2}(T)$ in both cases is
negative. This result suggests that the lattice geometry itself is 
insufficient
to explain the origin of the positive curvature of the upper critical field
observed in the copper--oxide high $T_c$ superconductors and additional
factors have to be taken into account. The most important feature of these
materials neglected in our approach is the presence of strong
correlations resulting
from the on--site repulsion on the copper sites, which can 
considerably suppress pairing in the $s$--wave channel. On the other hand,
in order to analyze the influence of these correlations on the upper critical
field one has to diagonalize the full Hamiltonian with the interaction term,
what is possible only for very small clusters for which the finite size
effects play much more important role. Apart from the finite size effects, 
there are too few points in the Brillouin zones of such small clusters
to reproduce the structure of the energy levels of infinite system.
In the framework of the proposed approach i.e., without the necessity 
of the above mentioned diagonalization of the Hamiltonian in the full Hilbert
space, one can take into account not only local, but also non--local pairing.
Such an extension would lead to an anisotropic superconductivity, which can
survive in spite of the presence of strong on--site repulsion 
\cite{Maki}. This problem
is currently under investigation.

\acknowledgments

Authors are grateful to Janusz Zieli{\'n}ski for a
fruitful discussion. This work has been supported by the
Polish State Committee for Scientific Research, Grant
No. 2 P03B 044 15.


\begin{thebibliography}{99}
\bibitem{gorkov} L. P. Gor'kov, Zh. Eksp. Teor. Fiz. {\bf 36}, 1918 (1959).
\bibitem{helf} We refer to: E. Helfand and N. R. Werthamer, Phys.
Rev. Lett. {\bf 13}, 686, (1964);
O. Fischer, Helv. Phys. Acta {\bf 45}, 332 (1972);
Y. N. Ovchinnikov and V. Z. Kresin
Phys. Rev. B {\bf 52}, 3075 (1995) for discussion of $H_{c2}$ in 3D and
2D free electron gas.
\bibitem{harper} P. G. Harper, Proc. Phys. Soc. London Sect. A {\bf 68}, 874 
(1955).
\bibitem{lang} D. Langbein, Phys. Rev. {\bf 180}, 633 (1969).
\bibitem{hof} R. D. Hofstadter, Phys. Rev. B {\bf 14} 2239 (1976). 
\bibitem{peierls} R. E. Peierls, Z. Phys. {\bf 80}, 763 (1933); 
J. M. Luttinger, Phys. Rev. {\bf 84}, 814 (1951). 
\bibitem{thoul} D. J. Thouless, in {\it The Quantum Hall Effect}, Graduate
Texts in Contemporary Physics (Springer--Verlag, New York, 1987), p. 101.
\bibitem{usov} N. A. Usov, Zh. Eksp. Teor. Fiz. {\bf 94}, 305 (1988). 
\bibitem{GL} A. A. Abrikosov, Zh. Eksp. Teor. Fiz. {\bf 32}, 1442 (1957).
\bibitem{LD} W. E. Lawrence and S. Doniach, in Proc. 12th Int. Conf. Low
Temp. Phys. Kyoto, edited by E. Kanda (Academic Press of Japan, Kyoto, 1971),
p. 361.
\bibitem{hc2} A. P. Mackenzie {\it et al.}, Phys. Rev. Lett. {\bf 71}, 1938
(1993); M. Osofski {\it et al.}, Phys. Rev. Lett. {\bf 71}, 2315 (1993). 
\bibitem{yoko} H. Yokoyama and H. Shiba, J. Phys. Soc. Jpn.
{\bf 56}, 1490 (1987).
\bibitem{tesan} M. Rasolt and Z. Tesanovi{\'c}, Rev. of Modern Phys.
{\bf 64}, 709 (1992).
\bibitem{Maki} We refer to: H. Won and K. Maki, Phys. Rev. B {\bf 53}, 5927
(1996); W. Kim, J. Zhu, and C. S. Ting, Phys. Rev. B {\bf 58}, R607
(1998) for theoretical discussion of $H_{c2}$ with anisotropic gap parameter.     
\end{thebibliography}
\end{document}